\documentclass[aps,prb,twocolumn,superscriptaddress]{revtex4-1} 
\usepackage{amsfonts}
\usepackage[dvips]{graphicx}
\usepackage{dcolumn}
\usepackage{bm}
\usepackage{amsmath}
\usepackage{amssymb}
\usepackage{revsymb4-1}
\usepackage{hhline}

\begin{document}

\title{Diamagnetism in wire medium metamaterials: theory and experiment}
\author{I. Yagupov}
\author{D. Filonov} 
\author {S. Kosulnikov}
\author {M. Hasan}
\author{I.V. Iorsh}
\author{P.A. Belov}

\affiliation{ITMO University, St. Petersburg 197101, Russia}

\begin{abstract}
Strong diamagnetic response of wire medium with finite wire radius is reported. Contrary to the previous works where it was assumed  that the wire medium exhibits only the electric response, we show   that  the non-zero magnetic susceptibility has to be taken into account for proper effective medium description of the wire medium. Analytical and numerical results are supported by the experimental measurements. 
\end{abstract}

\maketitle

\section{Introduction}
\begin{figure}[tbp]
\includegraphics[width= 0.65\columnwidth]{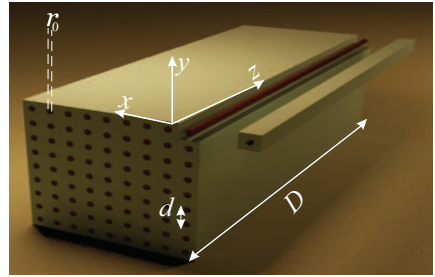} 
\caption{Geometry of the considered structure: array of perfectly conducting wires of finite radius $r_0$ and length $D$ is placed inside the dielectric matrix with permittivity $\varepsilon_m$.}
\label{FIGscheme}
\end{figure}

Wire metamaterials are composed of arrays of optically
thin metallic rods embedded in a dielectric matrix\cite{review}. These structures have been  extensively studied both theoretically and experimentally, due to  their possible applications in subwavelength image transfer\cite{imaging_1,imaging_2,imaging_3,imaging_4,imaging_5,imaging_6} and control over spontaneous emission lifetime of quantum emitters\cite{purcell_1,purcell_2}. While there are different geometries of the wire media, in our work we consider a special class, formed by a two-dimensional array of parallel wires of finite radius and length, shown schematically in Fig.~\ref{FIGscheme}.
For the practical applications and specifically in order to  model electromagnetic properties of wire metamaterial blocks of complex geometry, effective material parameters such as permittivity and permeability are required for proper homogenization. The effective dielectric permittivity of the array of parallel perfectly conducting wires has been obtained in a number of papers~\cite{hom0,hom01, Homogez1}. Permittivity and permeability tensors are given by~\cite{Homogez1}:
\begin{align}
\hat{\varepsilon}=\varepsilon_m\begin{pmatrix}
1 & 0 & 0 \\ 0 & 1 & 0 \\ 0 & 0 & \tilde{\varepsilon}_{zz}    
\end{pmatrix}, \quad \hat{\mu}=\mu_m, \mathrm{I}\label{epsprev}
\end{align} 
where $\mathrm{I}$ is the unity matrix,  $\varepsilon_m,\mu_m$ are the permittivity and permeability of the host material, and $\tilde{\varepsilon}_{zz}$ reads
\begin{align}
\tilde{\varepsilon}_{zz}=\left(1-\frac{2\left( r_0^2\log\frac{d^2}{4r_0(d-r_0)}\right)^{-1}}{(\omega/c)^2-k_z^2}\right),
\end{align}
where $r_0$ is the wire radius, and $d$ is the period of the structure. We note that the dielectric permittivity tensor is nonlocal, i.e. it depends on the wave vector $k_z$. However, in the case of an infinite wire structure along the lateral direction, at   normal incidence, we can consider only the tangential components of the dielectric permittivity which coincide with the dielectric permittivity of the matrix. Thus, at normal incidence, the light effectively interacts with uniform dielectric with permittivity $\varepsilon_m$ and magnetic permeability equal to unity. However, it is clear that the surface currents at the wires interfaces should lead to nonzero magnetic susceptibility. In this work, we present analytical, numerical, and experimental results to conclude that for the proper description of the wire media effective parameters it is required to account for the diamagnetic response of these structures.
\section{Theoretical modelling}
First, let us consider the quasistatic approximation for the single unit cell. We assume that the uniform static magnetic field is aligned along x direction. Then, the surface currents originating at the surface of the wire are aligned along the wires and can be expressed as $\mathbf{j}_{s,z}=-2\mu_m H \sin(\phi)$, where $\phi$ is the azimuthal angle. Knowing the currents we can calculate the magnetic moment of the unit cell $M$ as $\int d^3 \mathbf {r} [\mathbf{r}\times \mathbf{j}]$. Dividing it by the unit cell volume will give us the magnetization density and the effective permeability. The integral can be taken analytically yielding:
\begin{align}
\mu_{\mathrm{eff}}=\mu_m\left(1-\frac{4r_0^2}{d^2}\right). \label{mueff}
\end{align}
To obtain the effective permittivity we use the results of the work~\cite{Tyo}, where it is shown that wire medium belongs to a class of so-called $isorefractive$ media, where the relation $\varepsilon_{\mathrm{eff}}\mu_{\mathrm{eff}}=n^2=\varepsilon_{\mathrm{m}}\mu_m$ holds. Thus, for effective permittivity we have
\begin{align}
\varepsilon_{\mathrm{eff}}=\varepsilon_m \left(1-\frac{4r_0^2}{d^2}\right)^{-1} \label{epseff}.
\end{align}
Thus, we argue that for the correct description of the wire medium, $\varepsilon_m$ and $\mu_m$ in Eqs.~\eqref{epsprev} should substituted with right hand-sides of Eq.~\eqref{epseff} and Eq.~\eqref{mueff}, respectively. We note, that there are a lot of works dedicated to the artificial diamagnetism in metamaterials~\cite{Shamonina, Lapine, Diamat1}. At the same time, surprisingly, diamagnetism in wire media metamaterials is conventionally abandoned.

\begin{figure}[tbp]
\includegraphics[width= 0.98\columnwidth]{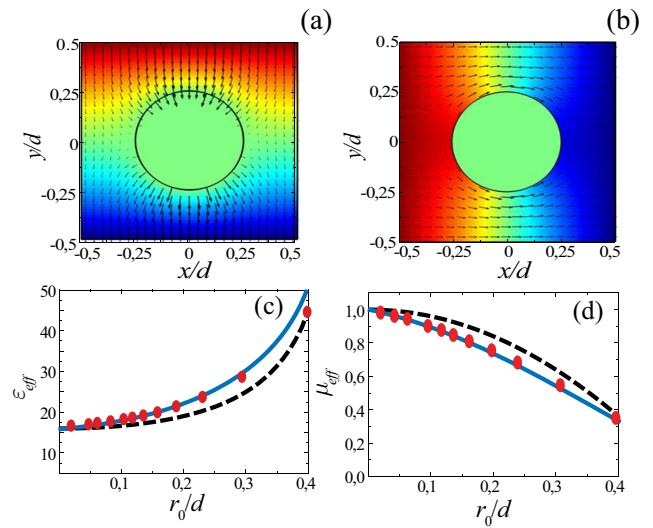} 
\caption{(a,b) Maps of the scalar electric (a) and magnetic (b)  potentials and field lines for the electric (a) and magnetic (b) fields in the unit cell for the case of $r_0/d=0.25$, the color scale is the same for the both maps. (c) Effective dielectric permittivity vs $r_0/d$ obtained via the analytical formula in Eq.~\eqref{epseff} (dashed black line), the semianalytical expression~\eqref{epssemi} (solid blue lines) and from extracting the effective parameters with NRW method (red circles). (d) Effective magnetic permeability vs $r_0/d$ obtained via the analytical formula~\eqref{mueff} (dashed black line), the semianalytical expression~\eqref{musemi} (solid blue lines) and from extracting the effective parameters  with NRW method (red circles)}. 
\label{figadd}
\end{figure}

We also support the analytical expressions with the numerical modelling. For that we use the homogenization technique similar to the one described in~\cite{SmithPendry}. We again assume the quasistatic approximation and consider the propagation of the TEM mode along the wires. In this case, we can set that electric field is aligned along $y$ and magnetic field along $x$ directions. We introduce static electric and magnetic potentials $u_e$, $u_m$ defined as $\bigtriangledown u_{e(m)}=\mathbf{E}(\mathbf{H})$. Then, two Laplace equations for  $u_e$ and $u_m$ can be solved. For the electric potential we use following boundary conditions: $u_e|_{y=\pm d/2}=\pm u_{e0}$, $\partial u_e /\partial x|_{x=\pm d/2} = 0 $, and $u_{e}|_{r=r_0}=0$ at the wire interface. For magnetic potential, boundary conditions are: $u_m|_{x=\pm d/2}=\pm u_{m0}$, $\partial u_m /\partial m|_{y=\pm d/2} = 0 $ and $\partial u_{\mathrm{m}}/\partial r|_{r=r_0} =0$. We solve the both Laplace equations numerically~\cite{Comsol}. Knowing the magnetic potentials and therefore the electric and magnetic fieds, it is possible to express the effective permittivity and magnetic permeability as
\begin{align}
\varepsilon_{\mathrm{eff}} =\frac{\int_{-d/2}^{d/2} \varepsilon_m E_y|_{y=-d/2} dx}{u_{e0}}, \label{epssemi}
\end{align}
\begin{align}
\mu_{\mathrm{eff}} =\frac{\int_{-d/2}^{d/2} \mu_m H_x|_{x=-d/2} dx}{u_{m0}}.\label{musemi}
\end{align}
The maps of the scalar electric and magnetic potentials as well as field lines are shown in Fig.\ref{figadd}(a,b). 
We compare the analytical expressions Eqs.~\eqref{mueff},\eqref{epseff} with the more rigorous approach given by expressions Eqs.~\eqref{epssemi},\eqref{musemi} in Figs.~\ref{figadd}(c,d), respectively. We use the following structure parameters: $\varepsilon_{\mathrm{m}}=16,\mu_{\mathrm{m}}=1$. It can be seen  that for thin wires, the diamagnetism is negligible and both approaches almost perfectly match. At the higher values of $r_0/d$ the two methods start to give slightly different results but still match quite well. The discrepancies at large $r_0/d$ appear, since the analytical approach considers a single wire and does not account for the structure periodicity and interaction between wires in different unit cells, which grows with the growth of $r_0/d$.

Moreover, in order to check if the quasistatic approximation is applicable in the case of realistic wire media structures, we have extracted the effective electric and magnetic susceptibility using Nicholas-Ross-Weir Method~\cite{NRW1,NRW2,NRW3}. We have calculated the reflection and transmission spectra of the wire medium slab with the same material parameters, thickness $D=10d$ in the full-wave electromagnetic  simulation package~\cite{CST}.The wire radius $r_0$ was a parameter which was changed in the region $[0.01d,0.4d]$. The conventional Nicolson-Ross-Weir (NRW) method  essentially is an inversion of the Rayleigh expressions for the transmission and reflection from the uniform material slab with respect to effective material refractive index $n=\sqrt{\varepsilon_{\mathrm{eff}}\mu_{\mathrm{eff}}}$ and impedance $Z=\sqrt{\varepsilon_{\mathrm{eff}}/\mu_{\mathrm{eff}}}$:
\begin{align}
&R=\frac{i(Z^2-1)}{2Z\cot(n\omega/c D)-i(Z^2+1)} \label{R_NRW},\\
&T=\frac{2Z/\sin(n\omega/c D)}{2Z\cot(n\omega/c D)-i(Z^2+1)}. \label{T_NRW}
\end{align}
 The values of $n$, $Z$ are  then substituted to  obtain the $\varepsilon_{\mathrm{eff}},\mu_{\mathrm{eff}}$. The results are shown with red dots in Figs.~\ref{figadd}(c,d). As shown in Figs.~\ref{figadd}(c,d) there is a good match between the full numerical simulation and semianalytical approach.

\begin{figure}[tbp]
\includegraphics[width= 0.95\columnwidth]{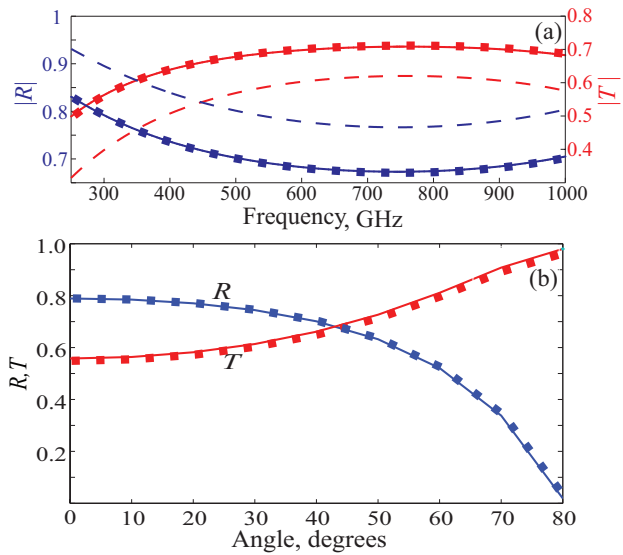} 
\caption{(a) reflection and transmission spectra for the case of normal incidence.  The solid and dashed lines correspond to the reflection and transmission calculated with Eqs.~\eqref{R_NRW},\eqref{T_NRW} with $\varepsilon=16,\mu=1$ (dashed lines), and with $\varepsilon=23.19,\mu=0.69$ (solid lines). The dots correspond to the CST modelling results. (b) The depedence of reflection and transmission on the incidence angle at the incident wave frequency $250$ GHz; solid lines correspond to  reflection and transmission calculated with Eqs.~\eqref{R_NRW},\eqref{T_NRW} with $\varepsilon=23.19,\mu=0.69$ and squares correspond to the CST modelling. Parameters of the wire medium structure are presented in the text.}
\label{FIGrt_freq}
\end{figure}

We stress that omitting the effective permeability of the structure leads to incorrect values of reflection and transmission coefficients if calculated with Rayleigh formulae, as shown in Fig.~\ref{FIGrt_freq}(a), where the reflection and transmission spectra for the cases when the permeability is accounted for or omitted are compared to the numerical modelling. The value of $r_0/d$ is set to 0.25. We can see that omitting the magnetic response gives the incorrect results for the reflection and transmission spectra.
Moreover, as can be seen in Fig.~\ref{FIGrt_freq}(b), the extracted parameters correctly describe the electromagnetic properties of the wire medium slab in the case of oblique incidence, when diamagnetism is taken into account. 
\section{Experimental measurements}
\begin{figure}[tbp]
\includegraphics[width= 0.95\columnwidth]{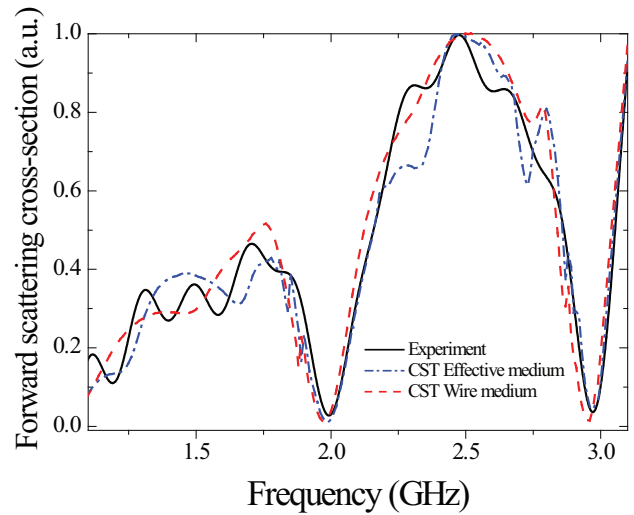} 
\caption{Scattering cross-section of the wire metamaterial block. The inset shows the photo of the structure. The experimentally determined scattering cross-section is compared with numerical calculation performed in CST. The ripples in the experiment are explained in the main text.}
\label{FIG_cs}
\end{figure}

In order to validate our results experimentally, we have constructed a wire metamaterial block with the following parameters: period of the structure $d=11$ mm, wire radius $r_0=5$ mm, wire length $D=150$ mm. The block consists of $20\times 20$ wires with air dielectric matrix, i.e.  $\varepsilon_m=1$. The photo of the structure is shown in the inset of Figure~\ref{FIG_cs} The experimental measurements were conducted in a following manner: a rectangular horn antenna (TRIM 0.75 GHz to 18 GHz; DR) connected to a transmitting port of the vector network analyzer Agilent E8362C was used to approximate a plane-wave excitation. The metamaterial block was  placed at the far-field region of the antenna  and a similar horn antenna (TRIM 0.75 GHz to 18 GHz) was employed as a receiver. The effective scattering cross-section   was obtained from the imaginary part of the forward scattering amplitude (due to the optical theorem). Due to the fact that the horn antenna modifies the wavefront substantially, we have normalized the  scattering cross-section  to unity. After completing the experimental measurements, we have performed the numerical simulation of the scattering at the metamaterial block of the same  dimensions for three different models. In the first model, the metamaterial block was modelled as an array of perfectly conducting wires, reconstructing the geometry of the experimental setup. In the second model, the block consisted of an uniform anistropic media with the following parameters: $\varepsilon_{\perp}=5.762,\varepsilon_{\|}=-10000.0+1.0i,\mu_{\perp}=0.1735,\mu_{\|}=1.0$, where subscripts $_{\|}$ and $_{\perp}$ correspond to the directions parallel and perpendicular to the wire axis, respectively. The scattering cross-sections were normalized  to unity. The above values for dielectric permittivities and magnetic permeabilities have been derived with the expressions in Eqs.~\eqref{mueff},~\eqref{epseff}. Finally in the third model,  we have calculated the scattering on metamaterial block made from the anisotropic uniform media with   the effective transverse dielectric and magnetic permeabilities equal to unity. Notably, it is quite obvious that in the third model, the scattering cross-section should be vanishingly small for the case of normal incidence, since both the effective impedance and refractive index of this structure are equal to unity.  The spectra of the forward scattering efficiency for the first two  cases are shown in Fig.~\ref{FIG_cs}. 
The observed oscillations in the experimental scattering spectra are due to the artificial Fabri-Perot resonances between metamaterial block interface and source and receiving horn antennas. 
We observe a good correspondence between the experimental cross-section spectra and numerical results, namely in the positions of the cross-section dips and in the cross-section modulation contrast. Even better correspondence is observed between the two numerical simulations. We note, that such correspondence could not be achieved in the approximation of effectively non-magnetic media, since as was discussed previously, such a media does not scatter the normally incident electromagnetic radiation at all. The observed discrepancies between the experimental and numerical simulation results are mainly due to the wavelength-dependent wave-front distortions of the incident wave, produced by the horn antenna. 
\section{Conclusion}
We have shown both theoretically and experimentally, that the thick wire media exhibit strong diamagnetic response, and that accounting for effective magnetic susceptibility is crucial for obtaining adequate material parameters of these structures.
\section{Acknowledgements}
We  thank A.N. Poddubny for helpful discussions. This work was supported by the Ministry of Education and Science of the Russian Federation (project 11.G34.31.0020, GOSZADANIE 2014/190, Zadanie No. 3.561.2014/K), by Russian Foundation for Basic Research and the Dynasty Foundation (Russia).


\begin{thebibliography}{99}
\bibitem{review} C. R. Simovski, P. A. Belov, A. V. Atrashchenko, and Y. S. Kivshar,
Adv. Mater. \textbf{24}, 4229 (2012).
\bibitem{imaging_1}P. A. Belov, Y. Zhao, S. Sudhakaran, A. Alomainy, and Y. Hao,
Appl. Phys. Lett. \textbf{89}, 262109 (2006).
\bibitem{imaging_2} A. Rahman, P. A. Belov, Y. Hao, and C. Parini, Opt. Lett. \textbf{35},
142 (2010).
\bibitem{imaging_3} G. Shvets, S. Trendafilov, J. B. Pendry, and A. Sarychev, Phys.
Rev. Lett. \textbf{99}, 053903 (2007).
\bibitem{imaging_4} S. Kawata, A. Ono, and P. Verma, Nature Photon. \textbf{2}, 438 (2008).
\bibitem{imaging_5} M. G. Silveirinha, P. A. Belov, and C. R. Simovski, Phys. Rev.
B \textbf{75}, 035108 (2007).
\bibitem{imaging_6} P. A. Belov, Y. Zhao, S. Tse, P. Ikonen, M. G. Silveirinha, C. R.
Simovski, S. Tretyakov, Y. Hao, and C. Parini, Phys. Rev. B \textbf{77},
193108 (2008).
\bibitem{purcell_1} A.N. Poddubny, P.A. Belov, Yu.S. Kivshar, Phys. Rev. B \textbf{87}, 035136  (2013).
\bibitem{purcell_2} A.P. Slobozhanyuk, A.N. Poddubny, A.E. Krasnok, and P.A. Belov,  Appl. Phys. Lett. \textbf{104}, 161105 (2014).
\bibitem{hom0} P. A. Belov, R. Marques,  S. I. Maslovski, I. S. Nefedov, M. Silveirinha, C. R. Simovski, and S. A. Tretyakov, Phys. Rev. B, \textbf{67}, 113103 (2003).
\bibitem{hom01} C.R. Simovski, P.A. Belov, Phys. Rev. E, \textbf{70}, 046616 (2004).
\bibitem{Homogez1} S. I. Maslovski and M. G. Silveirinha, Phys. Rev. B \textbf{80}, 245101
(2009).
\bibitem{Tyo} IEEE Trans.  Antennas and Propagation, \textbf{51}, 5, 1093 (2003).
\bibitem{Shamonina} E. Shamonina, L. Solymar, Eur. Phys. J. B \textbf{41}, 307–312 (2004).
\bibitem{Lapine}  M. Lapine, A. K. Krylova, P. A. Belov, C. G. Poulton, R. C. McPhedran, and Yu. S. Kivshar, Phys. Rev. B \textbf{87}, 024408 (2013).
\bibitem{Diamat1} L. Parke et al., Appl. Phys. Lett. \textbf{106}, 101908 (2015).
\bibitem{SmithPendry} D. R. Smith and J. B. Pendry, JOSA B, Vol. 23, Issue 3, pp. 391-403 (2006).
\bibitem{Comsol} The numerical modelling was performed in ComSol Multiphysics package.
\bibitem{NRW1} A. M. Nicolson and G. F. Ross, IEEE Trans. Instrum.
Meas., \textbf{19}, 4,  377  (1970).
\bibitem{NRW2} W. B. Weir, Proc. IEEE, \textbf{62}, 33 (1974).
\bibitem{NRW3} T. L. Blakney and W. B. Weir, Proc. IEEE, \textbf{63},  203 (1975).
\bibitem{CST} Simulations were performed in CST Microwave Studio Package.
\end{thebibliography}
\end{document}